\documentclass[portuguese,
  manuscript=article,  
  layout=publish,
  year=2024,
  volume=,
]{jmfis}

\usepackage[brazilian]{babel} 
\usepackage{amsmath}
\usepackage{dsfont}
\usepackage{bm}
\usepackage{cite}
\hypersetup{pdfborder = 0 0 0}

\doi{{ {\color{blue}\href{}{}}}}

\received {19/11/2024}
\accepted {19/11/2024}
\published{}

\newcommand{\orcidlink}[1]{\href{https://orcid.org/#1}{\includegraphics[width=10pt]{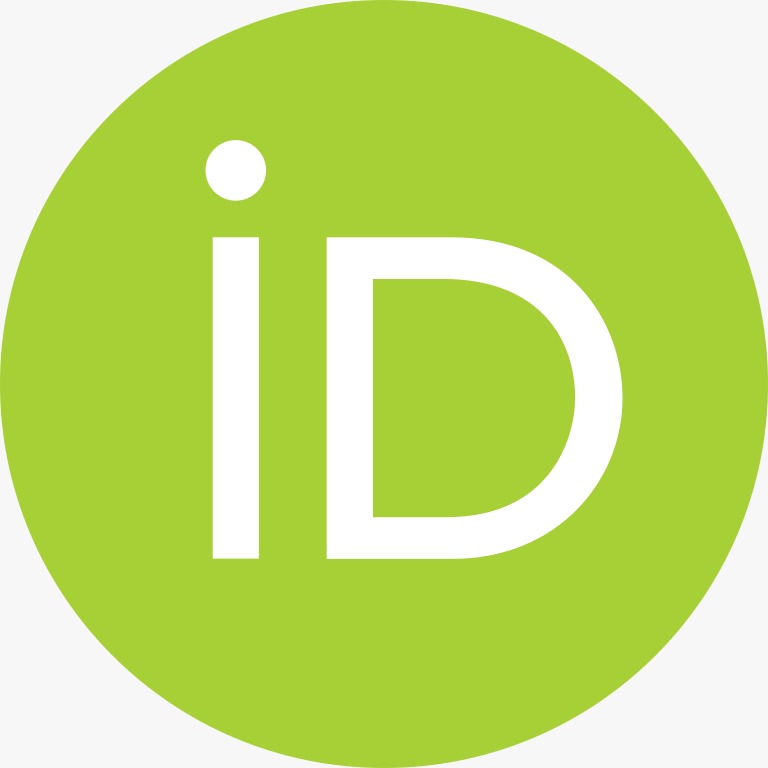}}}

\title{A Precisão da Metrologia Quântica: Limite de Cramér-Rao, Informação de Fisher e possíveis Aplicações Tecnológicas}

\author{Leonardo Antônio M. Souza \orcidlink{0000-0001-8374-984X}}
\email{leonardoamsouza@ufv.br}
\affiliation{Universidade Federal de Viçosa - Campus Florestal, 35690-000, Florestal, MG, Brasil}
\palavrachave{Metrologia; Metrologia Quântica; Estados Gaussianos}
\keywords{Metrology; Quantum Metrology; Gaussian States}

\begin{document}
\begin{resumo}
Este artigo explora o mais didaticamente possível os princípios fundamentais da metrologia, tanto clássica quanto quântica, com foco no Limite de Cramér-Rao e como ele define a precisão máxima na estimativa de parâmetros, levando em conta o ruído e a informação extraída dos dados. Fazemos um estudo pormenorizado também da Informação de Fisher (clássica e quântica), mostrando o significado Físico desta importante Figura de Mérito metrológica. Também discutimos como estados quânticos podem superar os limites clássicos, proporcionando uma precisão muito maior. Exemplos de aplicações tecnológicas incluem o desenvolvimento de sensores quânticos, termômetros quânticos e a estimativa de parâmetros de fase. Por fim, revisamos nossos resultados sobre a estimativa do parâmetro de compressão desconhecido aplicado num modo de um estado Gaussiano quântico.
\end{resumo}

\vspace{0.5cm}

\begin{abstract}
This paper explores as didactically as possible the fundamental principles of both classical and quantum metrology, focusing on the Cramér-Rao Bound and how it defines the maximum precision in parameter estimation, taking into account noise and the information extracted from the data. We also conduct a detailed study of Fisher Information (both classical and quantum), showing the physical significance of this important figure of merit in metrology. We further discuss how quantum states can surpass classical limits, providing much greater precision. Examples of technological applications include the development of quantum sensors, quantum thermometers, and phase parameter estimation. Finally, we review our results on the estimation of the unknown compression parameter applied to a mode of a quantum Gaussian state.
\end{abstract}

\section{Introdução}
Este artigo visa detalhar e contextualizar parte da apresentação de um dos autores na 20$^a$ Escola Matogrossense de Física (EMF), realizada em Outubro de 2024. Como em outro trabalho apresentado na 20$^a$ EMF tratamos do uso de RPG no Ensino de Física, iniciamos esta introdução de maneira lúdica.

Imagine Tyrion Lannister sendo encarregado de medir a altura do Trono de Ferro\cite{GoT}. O que inicialmente parece uma tarefa simples rapidamente se torna desafiador. O trono tem uma geometria complexa, com espadas de diferentes tamanhos projetando-se em todas as direções. Para conseguir uma medição precisa, Tyrion precisaria considerar todas essas variações de forma. Ele pode usar vários tipos de sistemas para medir a altura do trono, desde copos de vinho, palmos da mão do Rei, ou mesmo um tipo de equipamento chamado \emph{régua}. Mais que isso, para diminuir a incerteza com relação ao tamanho \emph{real} do Trono, Tyrion deve propor um tipo de estratégia que envolva a confecção de várias réguas aproximadamente iguais, de modo que os vários Lordes de Westeros possam medir o Trono.

Deixando de lado essa abordagem ``lúdica'' introdutória, nossa proposta nesse trabalho será tentar apresentar uma teoria para fazer a estimativa de um parâmetro desconhecido de um sistema. Podemos imaginar, pela história lúdica acima, que isso deve levar em conta:
\begin{itemize}
\item Estatística
\item Medidas
\item Parâmetro a ser estimado
\end{itemize}

Na área de \emph{Metrologia}, obter uma medida precisa de um sistema físico é um desafio que envolve muito mais do que uma simples régua - especialmente quando o sistema é quântico. Citamos por exemplo o Instituto Nacional de Metrologia, Qualidade e Tecnologia (INMETRO), que é a principal instituição brasileira responsável por garantir a padronização, a calibração e a qualidade das medições, assegurando que instrumentos e processos de medição no Brasil atendam a critérios de precisão e confiabilidade. Além disso, o INMETRO faz pesquisa de ponta em diversas áreas \cite{inmetro}.

Na física clássica, a precisão de uma estimativa está limitada pelo Limite de Cramér-Rao, que define a variância mínima de um estimador não enviesado, considerando o ruído e a quantidade de informação disponível nos dados. No entanto, a metrologia quântica oferece uma maneira de superar esse limite, utilizando propriedades quânticas como o emaranhamento. Neste artigo, exploramos como esses estados quânticos podem melhorar significativamente a precisão das medições e as implicações disso para a ciência e a tecnologia.

Como motivação para este trabalho, citamos que estados quânticos da Luz (estados comprimidos da Luz) têm sido utilizados para melhorar a precisão do sistema interferométrico usado no LIGO (Laser Interferometer Gravitational-Wave Observatory) \cite{LIGO}.

\begin{figure}
 \centering
 \includegraphics[scale=0.05]{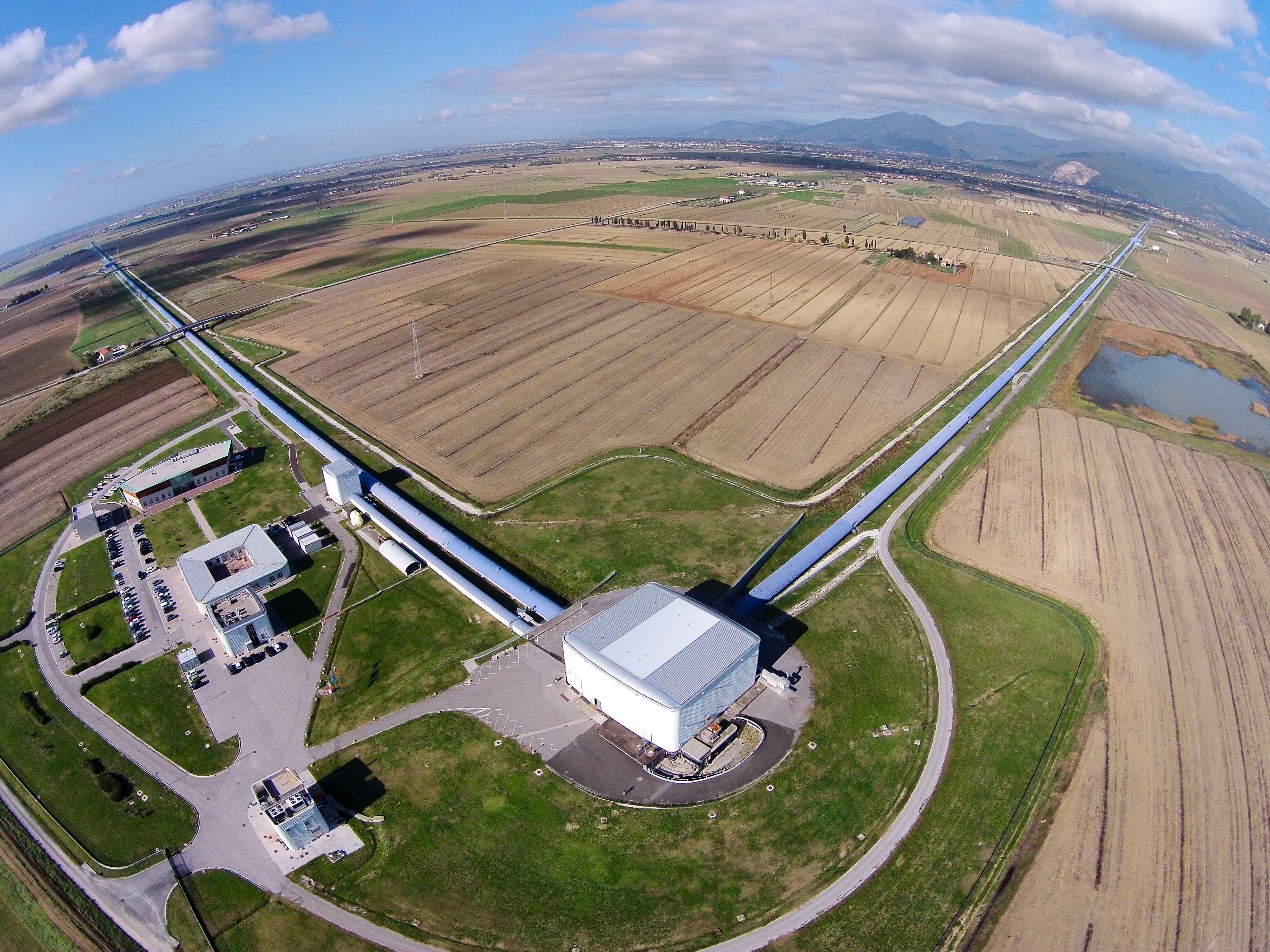}
 \caption{Vista aérea do LIGO (Laser Interferometer Gravitational-Wave Observatory), cada braço do interferômetro tem 4km de comprimento.}
 \label{figligo}
\end{figure}

Para além desta aplicação no LIGO, efeitos quânticos podem também melhorar a sensibilidade de sensores de gás \cite{sensor}, sensores automotivos \cite{cars}, processamento de imagens quânticos \cite{gabi}, termometria \cite{termometria}, sensores biomédicos \cite{biomedical} e relógios nucleares (que por sua vez poderão testar limites de teorias físicas) \cite{nuclear}.

Nossa proposta neste trabalho é apresentar, da maneira mais didática possível, uma teoria estatística,tanto  clássica quanto quântica, para estimar com a melhor precisão permitida o valor de um parâmetro desconhecido de um sistema. Este trabalho é dividido da seguinte forma: na seção \ref{classica} abordaremos o problema de estimativa de parâmetros clássica, na seção \ref{quantica} a teoria de estimativa de parâmetros quântica, na seção \ref{meu} apresentaremos alguns resultados de nosso grupo, e por fim concluiremos o trabalho na seção \ref{conclusao}

\section{Teoria de Estimativas Clássica}\label{classica}

\begin{figure}
 \centering
 \includegraphics[scale=0.23]{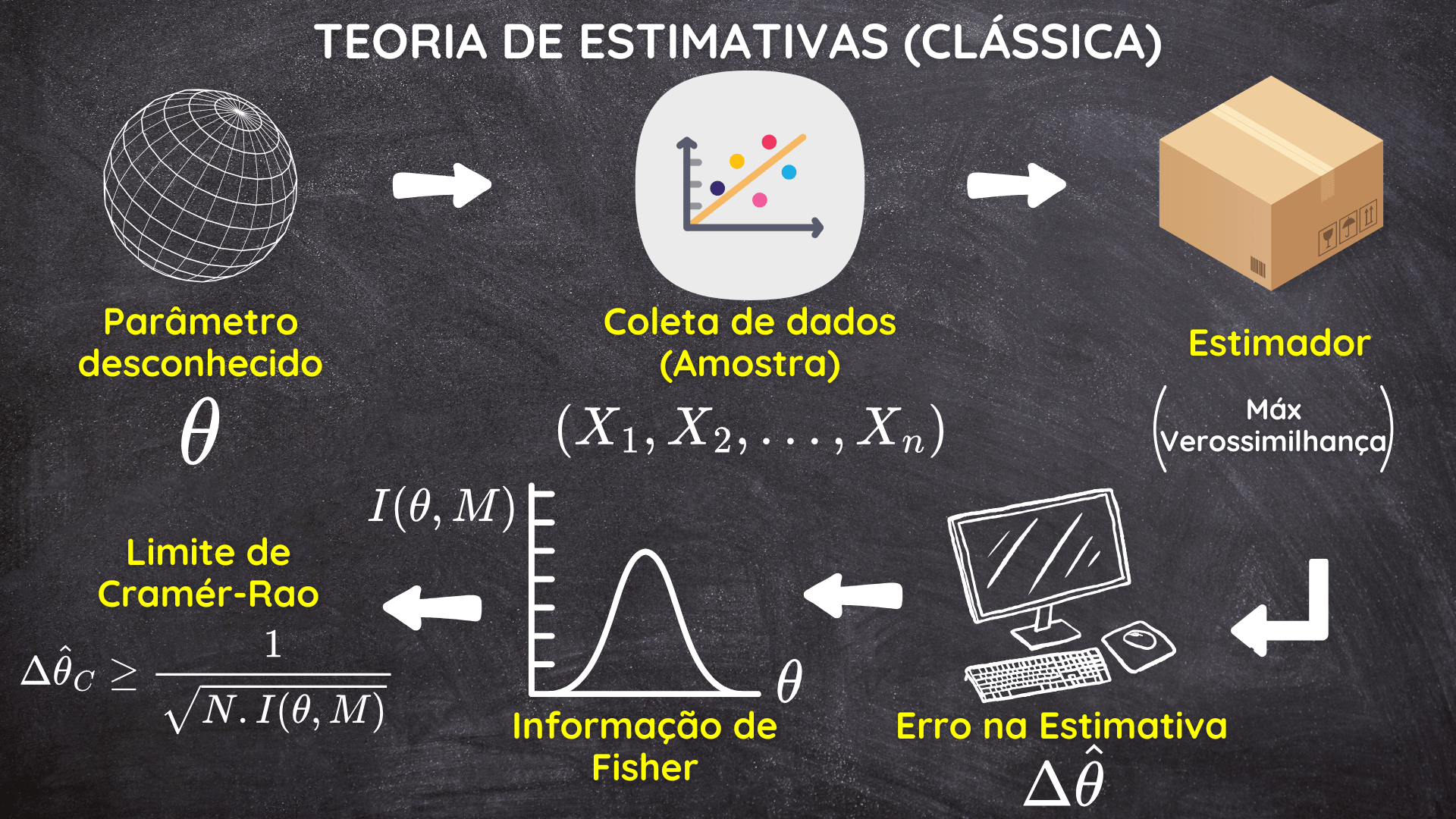}
 \caption{Uma típica estratégia para estimar um parâmetro $\theta$ desconhecido num sistema. Ver seção \ref{classica}.}
 \label{estrategia_classica}
\end{figure}

Em um típico problema de estimativa de parâmetros (ver Figura \ref{estrategia_classica}), um parâmetro desconhecido, por exemplo, $\theta$, é introduzido no sistema. Reúnem-se, então, os dados coletados em $N$ cópias desse sistema, por meio de medidas $M$ apropriadas, para construir um estimador\footnote{Em estatística, um estimador é uma regra para calcular uma estimativa de uma quantidade de interesse com base em dados observados, sendo distinguido do estimando (quantidade de interesse) e da estimativa (resultado obtido). A teoria da estimação estuda as propriedades dos estimadores para compará-los e determinar o melhor em diferentes situações, equilibrando robustez e precisão.}, que chamaremos de $\hat \theta$. Neste trabalho, não nos aprofundaremos no estudo do estimador e assumiremos que um estimador de máxima verossimilhança\footnote{O estimador de máxima verossimilhança (EMV) é um tipo específico de estimador que busca identificar o valor do parâmetro que maximiza a função de verossimilhança com base nos dados observados. Em outras palavras, enquanto o "estimador" é um termo genérico para qualquer regra que produz uma estimativa, o EMV é uma técnica específica que calcula essa estimativa selecionando o valor do parâmetro mais provável, dado o conjunto de dados disponíveis.} é suficiente para nossa apresentação.

A partir da construção do estimador e da análise dos dados coletados, podemos estimar o erro em nossa estimativa, representado por $\Delta \hat \theta$. Obviamente, queremos o menor erro possível, ou seja, que nosso estimador se aproxime do valor ``real'' do parâmetro $\theta$. Para estudar esse erro, podemos construir uma Figura de Mérito sensível a pequenas variações em torno de $\theta$, capaz de nos fornecer informações sobre o erro da estimativa. Essa Figura de Mérito será a Informação de Fisher, $I(\theta, M)$. O erro em nossa estimativa está relacionado à Informação de Fisher pelo Limite de Cramér-Rao:

\begin{equation} \Delta \hat \theta_{C} \geq \frac{1}{\sqrt{N \cdot I(\theta, M)}} \end{equation}

Uma de nossas tarefas neste artigo é apresentar didaticamente a Informação de Fisher e sua relação com a sensibilidade na estimativa de parâmetros. Em outras palavras, assumiremos que: há um parâmetro desconhecido $\theta$ que foi inserido no sistema; $N$ cópias desse sistema são medidas, e com esses dados construímos distribuições de probabilidades; e um estimador de máxima verossimilhança é construído. A partir desses pressupostos (que não serão detalhados neste artigo, mas recomendamos a leitura das referências \cite{metrology,livro,ballester,watanabe,leo} para mais detalhes), faremos um estudo da Informação de Fisher e sua relação com o erro na estimativa de parâmetros.

\subsection{O Limite de Cramér-Rao e a Estimativa de Parâmetros}
O Limite de Cramér-Rao é um conceito fundamental na teoria da estimação. Ele define o quão precisa pode ser a estimativa de um parâmetro em função da quantidade de ruído no sistema e da informação disponível nos dados. Matematicamente, o limite é dado por:
\begin{equation}
\Delta \hat{\theta}_C \geq \frac{1}{\sqrt{N \cdot I(\theta,M)}}
\end{equation}
Onde:
\begin{itemize}
 \item  $\hat{\theta}$ é o estimador não viesado do parâmetro $\theta$.
\item $I(\theta,M)$ é a informação de Fisher, que mede a quantidade de informação sobre o parâmetro $\theta$ contida nos dados.
\item $N$ é o número de cópias do sistema a ser avaliado.
\end{itemize}

O Limite de Cramér-Rao define a precisão máxima que se pode alcançar ao estimar um parâmetro, considerando o nível de ruído e a quantidade de informação disponível nos dados. Ele estabelece um limite inferior para a variância de qualquer estimador não viesado (unbiased), indicando quão precisa pode ser a estimativa de um parâmetro. Vamos tentar pensar sobre este limite através de um exemplo simples.

\subsubsection{Exemplo: Estimação da Posição de um Objeto com Ruído Gaussiano}\label{exemplo}

Considere o problema de estimar a posição verdadeira de um objeto, $\theta$, a partir de várias medições independentes (N medições). Em cada medição, existe um certo nível de ruído, assumido gaussiano e com variância conhecida $\sigma^2$.

Nosso objetivo é obter uma estimativa precisa para a posição verdadeira do objeto, $\theta$, com base nas medições ruidosas.

\textbf{Modelo de Medição}

Cada medição individual pode ser modelada como:
\begin{equation}
x_i = \theta + \text{ruído},
\end{equation}
onde o ruído é gaussiano com variância $\sigma^2$.

\textbf{Informação de Fisher}

Para esse caso específico, a Informação de Fisher é dada por (mostraremos como obter este resultado nas próximas seções):
\begin{equation}
I(\theta, M) = \frac{1}{\sigma^2}.
\end{equation}

Ao realizar N medições independentes, a Informação de Fisher acumulada se torna:
\begin{equation}
I_N(\theta, M) = \frac{N}{\sigma^2}.
\end{equation}

\textbf{Limite de Cramér-Rao}

O Limite de Cramér-Rao estabelece um limite inferior para a variância de qualquer estimador não viesado para $\theta$. Neste caso, esse limite é expresso como:
\begin{equation}
\Delta \hat{\theta}_C = \sqrt{\text{Var}(\hat{\theta})} \geq \frac{1}{\sqrt{I_N(\theta, M)}} = \frac{\sigma}{\sqrt{N}}.
\end{equation}

Esse limite representa a precisão máxima que se pode alcançar na estimativa de $\theta$, considerando o nível de ruído e o número de medições realizadas.

Sendo assim, o Limite de Cramér-Rao conclui que existe um limite fundamental de precisão (ou fronteira teórica) para a estimativa de um parâmetro físico, determinado pela quantidade de informação disponível nas medições. Observe que quanto menor o ruído $\sigma$, menor o erro. Note também que o erro $\Delta \hat \theta_C$ é inversamente proporcional à \emph{raiz quadrada de N}, isto será importante quando compararmos com o limite de Cramér-Rao quântico.

Mas e a Informação de Fisher? A Informação de Fisher está diretamente relacionada à quantidade de informação que um conjunto de dados contém sobre um sistema. Como estamos lidando com uma abordagem estatística, essa informação se encontra nas distribuições de probabilidade que descrevem cada conjunto de medições.

\subsection{Informação de Fisher (Clássica)}

Do que já mencionamos na seção \ref{classica}, é possível subentender que, dado um conjunto de dados extraídos das N cópias do sistema, obteremos distribuições de probabilidades, que nos retornarão quão próximos estamos do valor ``real'' do parâmetro em questão. Portanto, sejam $ p $ e $ q $ duas distribuições de probabilidade definidas sobre o mesmo espaço de saída $ \Omega $. A distância entre elas, conhecida como \emph{distância de Hellinger} \cite{metrology,livro,ballester,watanabe,leo}, é definida como:

\begin{equation}
d_H^2(q, p) = \sum_{\omega \in \Omega} \left( \sqrt{p(\omega)} - \sqrt{q(\omega)} \right)^2 = 2(1 - F(q, p)),
\end{equation}

onde $ F(q, p) = \sum_{\omega \in \Omega} \sqrt{p(\omega) q(\omega)} $ representa o \emph{overlap} ou sobreposição entre as distribuições $ p $ e $ q $. Esse overlap indica o grau de similaridade entre as duas distribuições: quanto maior o overlap, mais próximas estão as distribuições no sentido probabilístico.

Dada uma medida $ M $, consideramos que duas distribuições de probabilidade $ p(\theta, M) $ e $ p(\theta + \delta, M) $ são separadas por uma pequena quantidade $ \delta $, indicando que estamos interessados em pequenas variações do parâmetro $ \theta $. Nessa situação podemos fazer uma expansão da distância de Hellinger em torno de $\delta$, e assim temos:

\begin{equation}
d_H^2(p(\theta, M), p(\theta + \delta, M)) = \frac{1}{4} \sum_{\alpha, \beta} I(\theta, M)_{\alpha \beta} \, \delta_\alpha \delta_\beta + O(||\delta||^3),
\end{equation}

onde $ I(\theta, M)_{\alpha \beta} $ é um elemento da \emph{Matriz Informação de Fisher} \cite{metrology,livro,ballester,watanabe,leo}. A matriz Informação de Fisher mede a quantidade de informação que as distribuições fornecem sobre o parâmetro $ \theta $ e está definida como:

\begin{equation}
I(\theta, M)_{\alpha \beta} = \sum_{\omega \in \Omega^+} \frac{\partial_\alpha p(\theta, M) \, \partial_\beta p(\theta, M)}{p(\theta, M)},
\end{equation}

onde $ \partial_\alpha $ e $ \partial_\beta $ representam as derivadas parciais em relação aos parâmetros $ \alpha $ e $ \beta $, respectivamente. Aqui, $ \Omega^+ $ indica que a soma é realizada apenas sobre as saídas onde $ p(\omega) \neq 0 $, garantindo que a divisão seja bem definida.

No caso específico de uma estimativa uniparamétrica, ou seja, quando trabalhamos com um único parâmetro $ \theta $ (que é o caso tratado neste artigo), a Informação de Fisher simplifica-se para:

\begin{equation}
I(\theta, M) = \sum_{\omega \in \Omega^+} p(\theta, M) \left( \frac{\partial}{\partial \theta} \ln p(\theta, M) \right)^2.
\end{equation}


De forma geral, a Informação de Fisher fornece um indicador da sensibilidade do modelo estatístico a pequenas variações no parâmetro $ \theta $: quanto maior for o valor da informação de Fisher, mais precisas podem ser as estimativas do parâmetro, e portanto menor o erro da estimativa em questão.

Agora podemos entender o exemplo da seção \ref{exemplo}:

Considerando a função de densidade de probabilidade de uma distribuição normal (ruído Gaussiano com variância $\sigma^2$) dada por:

\begin{equation}
p(x, \theta, \sigma) = \frac{1}{\sqrt{2 \pi} \sigma} e^{-\frac{(x - \theta)^2}{2 \sigma^2}}.
\end{equation}

A Informação de Fisher é definida pela seguinte integral:

\begin{equation}
I(\theta, \sigma^2) = \int_{-\infty}^{\infty} p(x, \theta, \sigma^2) \left( \frac{\partial}{\partial \theta} \ln p(x, \theta, \sigma^2) \right)^2 \, dx.
\end{equation}

Calculamos a derivada do logaritmo da função de densidade em relação a $\theta$:

\begin{equation}
\frac{\partial}{\partial \theta} \ln p(x, \theta, \sigma) = \frac{(x - \theta)}{\sigma^2}.
\end{equation}

Substituindo essa derivada na expressão da Informação de Fisher, obtemos:

\begin{equation}
I(\theta, \sigma^2) = \int_{-\infty}^{\infty} p(x, \theta, \sigma) \left( \frac{x - \theta}{\sigma^2} \right)^2 \, dx.
\end{equation}

A integral pode ser calculada e, para a distribuição normal, o resultado que já apresentamos anteriormente é:

\begin{equation}
I(\theta, \sigma^2) = \frac{1}{\sigma^2}.
\end{equation}

Assim, concluímos que a Informação de Fisher para a distribuição normal fornece uma medida da quantidade de informação que os dados trazem sobre o parâmetro $\theta$, sendo inversamente proporcional à variância $\sigma^2$.

\subsection{Sumário: Teoria de Estimativas Clássica}

\begin{itemize}

\item A teoria de estimativas clássica estabelece um limite fundamental para a precisão de estimativas de parâmetros em sistemas com incertezas, conhecido como \emph{Limite de Cramér-Rao}. Esse limite é baseado na quantidade de informação que os dados contêm sobre o parâmetro de interesse, medida pela \emph{Informação de Fisher}.

\item O Limite de Cramér-Rao define um limite inferior para a variância de qualquer estimador não viesado de um parâmetro $ \theta $. Esse limite está diretamente relacionado à Informação de Fisher $ I(\theta, M) $, que mede a quantidade de informação disponível sobre o parâmetro $ \theta $ em uma dada medida $ M $. A desigualdade é dada por:

\begin{equation}
\text{Var}(\hat{\theta}) \geq \frac{1}{I(\theta, M)},
\end{equation}

onde $ \text{Var}(\hat{\theta}) = (\Delta \hat \theta)^2 $ representa a variância do estimador $ \hat{\theta} $. Esse limite implica que, quanto maior for a Informação de Fisher, menor será o erro associado à estimativa do parâmetro $ \theta $.

\item Para ilustrar o conceito de sensibilidade das distribuições de probabilidade em relação ao parâmetro $ \theta $, introduzimos a \emph{Distância de Hellinger}. A distância de Hellinger entre duas distribuições de probabilidade $ p $ e $ q $ sobre o mesmo espaço de saída $ \Omega $ é definida como:

\begin{equation}
d_H^2(q, p) = \sum_{\omega \in \Omega} \left( \sqrt{p(\omega)} - \sqrt{q(\omega)} \right)^2 = 2(1 - F(q, p)),
\end{equation}

onde $ F(q, p) = \sum_{\omega \in \Omega} \sqrt{p(\omega) q(\omega)} $ é o \emph{overlap} entre as distribuições $ p $ e $ q $.

\item A Informação de Fisher está associada ao comportamento da distância de Hellinger em pequenas variações do parâmetro $ \theta $. Quando duas distribuições $ p(\theta, M) $ e $ p(\theta + \delta, M) $ são separadas por uma pequena variação $ \delta $, a distância de Hellinger entre elas é dada por:

$$
d_H^2(p(\theta, M), p(\theta + \delta, M)) = \frac{1}{4} \sum_{\alpha, \beta} I(\theta, M)_{\alpha \beta} \, \delta_\alpha \delta_\beta + O(||\delta||^3),
$$

onde $ I(\theta, M)_{\alpha \beta} $ representa um elemento da matriz de Informação de Fisher. Esse resultado mostra que, quanto maior for a Informação de Fisher, maior será a distância entre as distribuições de probabilidade associadas a diferentes valores do parâmetro $ \theta $.

\item Em resumo, a Informação de Fisher quantifica a sensibilidade das distribuições de probabilidade em relação a variações do parâmetro $ \theta $. Especificamente:
\begin{itemize}
    \item Quanto maior a Informação de Fisher, menor será o erro associado à estimativa do parâmetro, conforme estabelecido pelo Limite de Cramér-Rao.
    \item Quanto maior a Informação de Fisher, maior será a distância entre distribuições de probabilidade próximas, o que indica que a distribuição é mais sensível a pequenas mudanças no parâmetro.
\end{itemize}

\item Assim, a Informação de Fisher serve como uma medida da precisão potencial em estimativas de parâmetros, refletindo a capacidade dos dados de distinguir diferentes valores de $ \theta $.

\end{itemize}

\section{Postulados da Teoria Quântica}\label{postulados}

Por completeza, apresentamos nesta breve seção os Postulados da Mecânica Quântica, em sua formulação tradicional (ver \cite{griffiths, cohen, messiah, nielsen}).

Os postulados da mecânica quântica, considerando tanto a abordagem por vetores de estado quanto a formulação de matriz densidade, podem ser expressos da seguinte maneira: a informação sobre o sistema quântico é representada por um vetor de estado $ |\psi\rangle $ em um espaço de Hilbert na abordagem de kets e bras, e por uma matriz densidade $ \rho $ na formulação de matriz densidade; os observáveis são correspondentes a operadores hermitianos $ \hat{A} $ que atuam sobre este espaço, onde a probabilidade de obter um autovalor específico $ a_n $ ao realizar uma medição é dada por:

$$
P(a_n) = |\langle \phi_n | \psi \rangle|^2,
$$

na abordagem por vetor de estado, e por

$$
P(a_n) = \text{Tr}(\rho P_n),
$$

na abordagem de matriz densidade, onde $ P_n = |\phi_n\rangle \langle \phi_n| $ é o projetor associado ao autovetor correspondente; a evolução do estado quântico do sistema é descrita pela equação de Schrödinger, expressa como:

$$
i \hbar \frac{\partial}{\partial t} |\psi(t)\rangle = \hat{H} |\psi(t)\rangle,
$$

que se traduz na evolução da matriz densidade, para evoluções unitárias, como (equação de Liouville-von Neumann):

$$
\frac{\partial \rho}{\partial t} = -\frac{i}{\hbar} [\hat{H}, \rho],
$$

onde $[ \hat{H}, \rho ]$ é o comutador entre o Hamiltoniano e a matriz densidade; finalmente, o processo de medida colapsa o vetor de estado em um dos autovetores do operador observável, resultando em um valor mensurável $ a_n $, descrito por:

$$
\rho' = \frac{P_n \rho P_n}{\text{Tr}(P_n \rho P_n)},
$$

na abordagem de matriz densidade, enquanto na abordagem de vetores de estado, o colapso do estado é simplesmente dado por:

$$
|\psi\rangle \rightarrow |\phi_n\rangle,
$$

onde $ |\phi_n\rangle $ é o autovetor associado ao autovalor medido $ a_n $.

\section{Teoria de Estimativas Quântica}\label{quantica}

\begin{figure}
 \centering
 \includegraphics[scale=0.23]{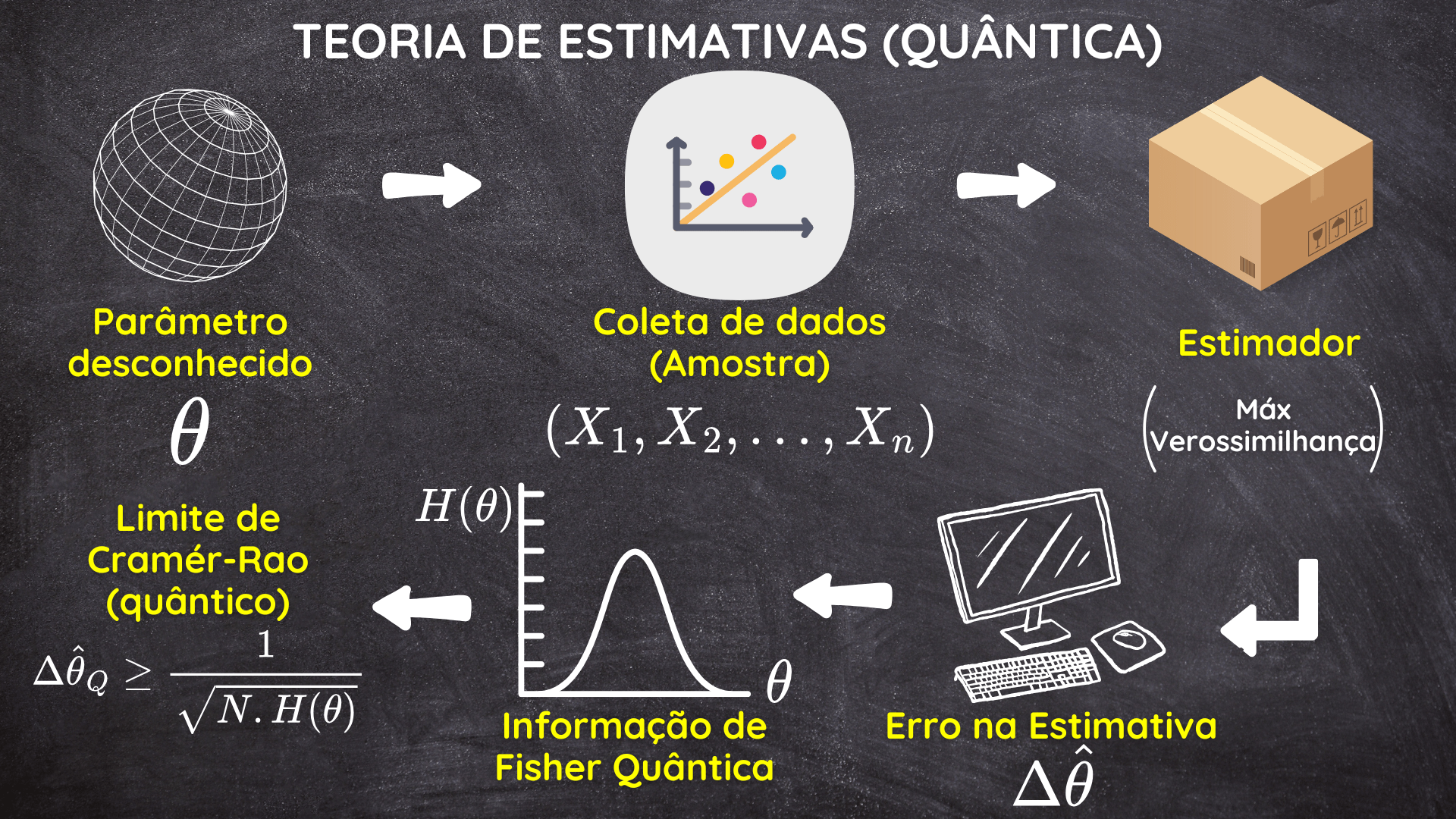}
 \caption{Uma típica estratégia para estimar um parâmetro $\theta$ desconhecido num sistema. Ver seção \ref{quantica}. Observe que nesta estratégia, a Figura de Mérito $H(\theta)$, a Informação de Fisher Quântica, não depende das medidas $M$.}
 \label{estrategia_quantica}
\end{figure}

Assim como numa estratégia de estimativa clássica, uma estratégia quântica terá passos similares, porém podemos ter acesso a recursos quânticos como estados de superposição ou Emaranhamento, e também, como visto na seção \ref{postulados}, a medida possui um caráter definidor quando tratamos de mecânica quântica. Sendo assim, um típico problema de estimativa de parâmetros usando recursos quânticos (ver Figura \ref{estrategia_quantica}), um parâmetro desconhecido, por exemplo, $\theta$, é introduzido no sistema. Este sistema pode possuir características quânticas, que podem ser usadas para melhorar a estimativa do parâmetro em questão. Reúnem-se, então, os dados coletados em $N$ cópias desse sistema, por meio de medidas $M$ apropriadas, para construir um estimador $\hat \theta$. Novamente assumiremos um estimador de máxima verossimilhança.

A sequência é similar: da obtenção da Informação de Fisher Quântica $H(\theta)$ que, diferentemente da Informação de Fisher Clássica não depende das medidas $M$, estudar o erro $\Delta \hat \theta_Q$, através da versão quântica do limite de Cramér-Rao.

\subsection{Informação de Fisher (Quântica)}

No caso quântico, para obtermos a Informação de Fisher, devemos iniciar nossa análise não com as distribuições de probabilidade, mas avaliando a distância entre dois estados quânticos. Sejam então $ \rho_1 $ e $ \rho_2 $ dois estados quânticos que possam gerar duas distribuições de probabilidades $p(\rho_1,M)$ e $p(\rho_2,M)$ sobre o mesmo espaço de saída $ \Omega $. Definimos a \emph{distância de Bures} \cite{metrology,livro,ballester,watanabe,leo} como:

\begin{equation}
 d_B^2 (\rho_1, \rho_2) = \max_M d_H^2 [p(\rho_1,M),p(\rho_2,M)] = 2 ( 1-\sqrt{F(\rho_1,\rho_2)} ),
\end{equation}
com $F(\rho_1,\rho_2) = \text{Tr} \sqrt{\sqrt{\rho_1}\rho_2\sqrt{\rho_1}}$ a Fidelidade entre os estados. Observe que, ao realizarmos uma maximização em todas as Medidas $M$, já fazemos que a distância de Bures independa da medida em si.

Se queremos saber se os dois estados quânticos são ``próximos'', podemos avaliar esta distância quando $\rho_1$ e $\rho_2$ diferem de uma quantidade pequena $\delta$, e posteriormente fazer uma expansão, encontrando:

\begin{equation}\label{bures}
 d_B^2 (\rho_1,\rho_2) = \frac{1}{4} H(\theta)  \delta^2 + O(\delta^3).
\end{equation}

Na equação \eqref{bures}, a função $H(\theta)$, que depende somente do parâmetro a ser estimado, é dada por \cite{ballester, leo}:

\begin{equation}\label{informacao_fisher}
H(\theta) = Tr[\rho(\theta) L_\theta^2 ],
\end{equation}

onde $L_\theta$ é a chamada \emph{derivada logarítmica simétrica}, e é definida a partir da seguinte equação:

\begin{equation}\label{SLD}
\frac{\partial \rho(\theta)}{\partial \theta} = \frac{1}{2} \left( L_\theta \rho(\theta) + \rho(\theta) L_\theta \right).
\end{equation}

Assim como no caso clássico, podemos dar a seguinte interpretação para a Informação de Fisher:
\begin{itemize}
 \item Quanto maior a Informação de Fisher Quântica, maior a sensibilidade do estado quântico a pequenas mudanças em $\theta$, o que aumenta a ``distância'' ou a diferença entre esses estados quânticos no espaço de Hilbert, refletida pela maior separação medida pela Distância de Bures. A Informação de Fisher quântica captura a sensibilidade dos estados quânticos a mudanças no parâmetro $\theta$, refletida pela curvatura da Distância de Bures.
\end{itemize}

Uma breve observação sobre a derivada logarítmica simétrica \eqref{SLD}: $L_\theta$ é um operador; ao encontrarmos os autovetores da derivada logarítmica simétrica, eles formam o melhor conjunto de POVM para o problema em questão.

\subsection{Limite de Cramér-Rao Quântico}

Uma propriedade importante da informação de Fisher quântica (IFQ) é a aditividade para sistemas independentes. Considere que temos \( N \) cópias de um sistema quântico, e seja \( H^N(\theta) \) a informação de Fisher global associada ao parâmetro \(\theta\) para o sistema total. Nesse caso, a IFQ global é simplesmente a soma das informações de Fisher de cada cópia, levando à expressão:

\[
H^N(\theta) = N \, H(\theta),
\]

onde \( H(\theta) \) representa a informação de Fisher quântica para uma única cópia do sistema.

Um resultado fundamental associado à IFQ é a chamada \emph{desigualdade de Braunstein-Caves}, que estabelece um limite para a precisão da estimação de parâmetros em sistemas quânticos. Seja \(\rho\) um estado quântico dependente de um parâmetro \(\theta\), e \( M \) uma medida arbitrária aplicada ao sistema. A desigualdade de Braunstein-Caves afirma que a informação de Fisher clássica \( I_M(\theta) \), associada à medida \( M \), é limitada pela informação de Fisher quântica:

\[
I(\theta,M) \leq H(\theta).
\]

Esse limite garante que, para qualquer escolha de medida, a informação acessível sobre o parâmetro \(\theta\) está sempre restrita pelo valor da IFQ, que representa o limite máximo de informação disponível em um sistema quântico sobre o parâmetro \(\theta\).

A partir dessa desigualdade, segue o chamado limite de Cramér-Rao quântico. Considere um estimador não tendencioso \(\hat{\theta}\) do parâmetro \(\theta\). Sob certas condições (que não exploraremos em detalhes aqui), a variância desse estimador satisfaz a seguinte desigualdade:

\[
\Delta \hat \theta \geq \frac{1}{\sqrt{H(\theta)}},
\]

onde $\Delta \hat \theta$ representa a raiz da variância do estimador \(\hat{\theta}\) em relação ao parâmetro verdadeiro \(\theta\) e à medida \(M\). Esse limite estabelece que a precisão com que \(\theta\) pode ser estimado é limitada pela inversa da informação de Fisher quântica \( H(\theta) \), definindo, assim, um limite inferior para a variância de qualquer estimador não enviesado do parâmetro \(\theta\).

\subsection{Limite padrão quântico (Shot noise limit) e Limite de Heisenberg}

A partir da informação de Fisher quântica e da desigualdade de Cramér-Rao quântico, podemos compreender dois limites fundamentais na metrologia quântica: o \emph{standard quantum limit} (SQL) e o limite de Heisenberg.

O \emph{standard quantum limit} estabelece que, ao se utilizar \( N \) cópias independentes de um sistema quântico para estimar um parâmetro \(\theta\), o erro na estimativa, definido como a raiz quadrada da variância do estimador \(\hat{\theta}\), é limitado pela ordem de \( N^{-1/2} \). Ou seja, temos:

\[
\Delta \theta \geq \frac{1}{\sqrt{N}},
\]

onde \( \Delta \theta\) representa o erro padrão na estimativa de \(\theta\). Esse limite é atingido em medições independentes e representa a precisão máxima possível sem o uso de correlações quânticas entre as cópias.

Por outro lado, o \emph{limite de Heisenberg} define um limite ainda mais restrito quando se faz uso de emaranhamento ou outras correlações quânticas entre as cópias. Nesse caso, o erro padrão pode, em princípio, atingir a ordem de \( N^{-1} \):

\[
\Delta \theta \geq \frac{1}{N}.
\]

Esse limite é mais preciso que o SQL e representa a máxima precisão teoricamente alcançável para a estimativa de \(\theta\) ao se explorar totalmente os recursos quânticos do sistema.

Outro resultado \cite{escher} importante nos diz que, ao trabalharmos com a Energia do estado quântico em questão, em especial ao tratarmos de estados Gaussianos (ver seção \ref{meu}), o limite de Heisenberg pode ser alcançado quando: $$H(\theta) \sim n^2,$$ onde $n$ é o número de fótons do modo Gaussiano estudado.

Deixamos aqui uma observação: o uso de recursos quânticos (estados de Fock, estados de superposição, Emaranhamento, etc.) permitem que, em princípio e em muitos casos, possamos saturar o limite de Heisenberg na estimativa de parâmetros\cite{classes, Luca}.

\subsection{Sumário: Teoria de Estimativas Quânticas}

\begin{itemize}

\item A teoria de estimativas quânticas expande a teoria clássica, levando em consideração as propriedades específicas dos sistemas quânticos, como o princípio da incerteza e o emaranhamento. A precisão com que podemos estimar um parâmetro \( \theta \) em um sistema quântico é limitada pela \emph{Informação de Fisher Quântica} (IFQ) e pelo \emph{Limite de Cramér-Rao Quântico} (QCRB).

\item O \emph{Limite de Cramér-Rao Quântico} (QCRB) estabelece um limite inferior para o erro da estimativa de um parâmetro \( \theta \) em sistemas quânticos. Esse limite é inversamente proporcional à Informação de Fisher Quântica (\( H_\theta \)), que quantifica a quantidade de informação sobre o parâmetro presente no estado quântico do sistema. A desigualdade é dada por:

\begin{equation}
(\Delta \hat{\theta})^2 \geq \frac{1}{H_\theta},
\end{equation}

onde \( (\Delta \hat{\theta})^2 \) representa o erro quadrático médio do estimador \( \hat{\theta} \), e \( H_\theta \) é a Informação de Fisher Quântica associada ao estado quântico e ao operador de Hamiltoniano que descreve a dependência do sistema em relação ao parâmetro \( \theta \).

\item A \emph{Informação de Fisher Quântica} (IFQ) quantifica a capacidade de um estado quântico de fornecer informações sobre o parâmetro \( \theta \). A IFQ é dada pela expressão:

\begin{equation}
H_\theta = \text{Tr} \left( \rho  L_\theta^2 \right),
\end{equation}

onde \( \rho \) é o operador densidade do estado quântico e \( L_\theta \) é a derivada logarítmica simétrica.

\item A IFQ está diretamente relacionada à precisão da estimativa de parâmetros em sistemas quânticos. Quanto maior for a IFQ, mais precisa será a estimativa do parâmetro \( \theta \), seguindo o Limite de Cramér-Rao Quântico.

\item Em sistemas quânticos compostos, como os de múltiplos modos, a precisão da estimativa de parâmetros pode ser aprimorada com o uso de estados comprimidos ou emaranhados. A IFQ pode ser generalizada para sistemas de dois ou mais modos, levando em conta as interações quânticas entre os modos.

\item O \emph{Limite de Cramér-Rao Quântico} (QCRB) impõe uma limitação fundamental para a precisão de qualquer estimativa não viesada em sistemas quânticos. Esse limite depende da IFQ e reflete o impacto das propriedades quânticas do sistema, como a coerência e o emaranhamento, na precisão das medições.

\item A presença de \emph{emaranhamento quântico} entre os modos de um sistema pode melhorar a precisão das estimativas de parâmetros. Estados emaranhados aumentam a IFQ, permitindo que os sistemas quânticos ultrapassem os limites de precisão impostos pela teoria clássica.

\item A teoria de estimativas quânticas é uma ferramenta central na \emph{metrologia quântica}, que visa explorar os recursos quânticos para melhorar a precisão das medições. A metrologia quântica tem aplicações em tecnologias de sensores quânticos, interferometria quântica e experimentos de alta precisão, como no caso dos detectores de ondas gravitacionais.

\item Em resumo, a teoria de estimativas quânticas fornece os fundamentos para entender e explorar a precisão das medições quânticas. Especificamente:

\begin{itemize}
    \item O Limite de Cramér-Rao Quântico (QCRB) estabelece limites fundamentais para a precisão de estimativas em sistemas quânticos, considerando os efeitos da mecânica quântica.
    \item A Informação de Fisher Quântica (IFQ) quantifica a quantidade de informação disponível sobre o parâmetro \( \theta \) em um estado quântico.
    \item Estados com propriedades intrinsicamente quânticas (como emaranhamento) podem melhorar a precisão das estimativas, permitindo que sistemas quânticos ultrapassem as limitações clássicas de precisão.
\end{itemize}

\end{itemize}

\section{Estimativa do Parâmetro de Compressão num estado Gaussiano}\label{meu}

\subsection{Estados Gaussianos numa casca de noz}\label{sec_gaussian}

Aqui vamos apresentar rapidamente alguns resultados sobre estados Gaussianos de um ou dois modos submetidos a dinâmicas Gaussianas, que preservam a forma Gaussiana do estado \cite{adesso2007entanglement}. Um sistema de variáveis contínuas (CV) de dois modos \( A \) e \( B \), com operadores de aniquilação \( \hat{a} \) e \( \hat{b} \), é descrito pelo vetor de quadratura \( \hat{\boldsymbol{O}} = \{ \hat{q}_A, \hat{p}_A, \hat{q}_B, \hat{p}_B \} \), e satisfaz relações de comutação canônica com a forma simplética.

Estados Gaussianos \( \rho_{AB} \) \cite{adesso2007} são caracterizados pelo vetor de deslocamento \( \boldsymbol{\varepsilon}_{AB} \) e pela matriz de covariância \( \boldsymbol{\sigma}_{AB} \), que satisfazem a relação de incerteza de Robertson-Schrödinger, dada por:

\[
\boldsymbol{\sigma}_{AB} + i \boldsymbol{\Omega} \geq 0,
\]

onde \( \boldsymbol{\Omega} \) é a forma simplética associada ao sistema, dada por:

\[
\boldsymbol{\Omega} =
\left(
\begin{array}{cc}
0 & 1 \\
-1 & 0
\end{array}
\right)^{\oplus 2}.
\]

A matriz de covariância para um estado de dois modos pode ser escrita de maneira geral como:

\[
\boldsymbol{\sigma}_{AB} =
\left(
\begin{array}{cccc}
a_1 & g & c & 0 \\
g & b_1 & 0 & d \\
c & 0 & b_2 & 0 \\
0 & d & 0 & b_2
\end{array}
\right),
\]

onde os coeficientes são definidos de forma que a matriz \( \boldsymbol{\sigma}_{AB} \) satisfaça a condição física da relação de incerteza de Robertson-Schrödinger. Essa matriz pode ser transformada, por meio de operações simpléticas locais, para uma forma padrão, com sub-blocos diagonais \( 2 \times 2 \), dada por:

\[
\boldsymbol{\sigma}_{AB} =
\left(
\begin{array}{cc}
\boldsymbol{\alpha} & \boldsymbol{\gamma} \\
\boldsymbol{\gamma}^T & \boldsymbol{\beta}
\end{array}
\right),
\]

onde \( \boldsymbol{\alpha} = \text{diag}\{a, a\} \), \( \boldsymbol{\beta} = \text{diag}\{b, b\} \) e \( \boldsymbol{\gamma} = \text{diag}\{c, d\} \), com as condições \( a, b \geq 1 \) e \( c \geq |d| \geq 0 \).

Definimos o número médio total de excitações como \( E \equiv \bar{n}_A + \bar{n}_B = 2 \bar{n} \), onde \( \bar{n}_A \) e \( \bar{n}_B \) são os números médios de excitações nos modos \( A \) e \( B \), respectivamente. Eles são dados por:

\[
\bar{n}_A = \frac{\text{Tr}[\boldsymbol{\alpha}] - 2}{4} + \frac{\varepsilon_{x,A}^2 + \varepsilon_{p,A}^2}{4}, \quad \bar{n}_B = \frac{\text{Tr}[\boldsymbol{\beta}] - 2}{4} + \frac{\varepsilon_{x,B}^2 + \varepsilon_{p,B}^2}{4},
\]

onde \( \boldsymbol{\alpha} \) e \( \boldsymbol{\beta} \) são as submatrizes diagonais da matriz de covariância associada aos modos \( A \) e \( B \), e \( \varepsilon_{x,A}, \varepsilon_{p,A}, \varepsilon_{x,B}, \varepsilon_{p,B} \) são os componentes do vetor de deslocamento \( \boldsymbol{\varepsilon} \).

Para estados compostos, utilizamos a \emph{Negatividade Logarítmica}, \( \mathcal{E}_N \), como medida de emaranhamento. A Negatividade Logarítmica é dada por:

\[
\mathcal{E}_N = \max \left( 0, - \ln \tilde{\nu} \right),
\]

onde \( \tilde{\nu} \) é o menor autovalor simplético da transposta parcial da matriz de covariância. Os autovalores simpléticos \( \tilde{\nu} \) podem ser calculados a partir dos invariantes simpléticos \( A \), \( B \), \( C \), e \( D \) da matriz de covariância \( \boldsymbol{\sigma}_{AB} \), e são dados pela expressão:

\[
2 \tilde{\nu}^2 = H - \sqrt{H^2 - 4D},
\]

onde \( H = A + B - 2C \).

Além disso, introduzimos uma medida de \emph{Coerência} para analisar estados de um único modo, com base na entropia do estado \( S(\rho) \) e no número médio de fótons \( \bar{n}_i \) \cite{xu2016quantifying}. A coerência é dada por:

\[
C(\rho) = - S(\rho) + \sum_{i=1}^N \left[ (\bar{n_i}+1) \log_2 (\bar{n_i}+1) - \bar{n_i} \log_2 \bar{n_i} \right],
\]

onde \( S(\rho) \) é a entropia do estado \( \rho \) e \( \bar{n}_i \) é o número médio de fótons do \( i \)-ésimo estado. A entropia \( S(\rho) \) é dada por:

\[
S(\rho) = - \sum_{i=1}^N \left[ \left( \frac{\nu_i - 1}{2} \right) \log_2 \left( \frac{\nu_i - 1}{2} \right) - \left( \frac{\nu_i + 1}{2} \right) \log_2 \left( \frac{\nu_i + 1}{2} \right) \right],
\]

onde \( \{ \nu_i \}_{i=1}^N \) são os autovalores simpléticos da matriz de covariância. Neste contexto, estados puros são os que apresentam máxima coerência, e é intuitivo observar que estados comprimidos podem ter valores de coerência maiores do que os estados coerentes convencionais.

\subsection{Nossa estratégia para estimar o parâmetro de compressão num Estado Gaussiano}\label{estrategia}

\begin{figure}[h]
 \begin{center}
 \hspace*{-1cm}\includegraphics[scale=0.33]{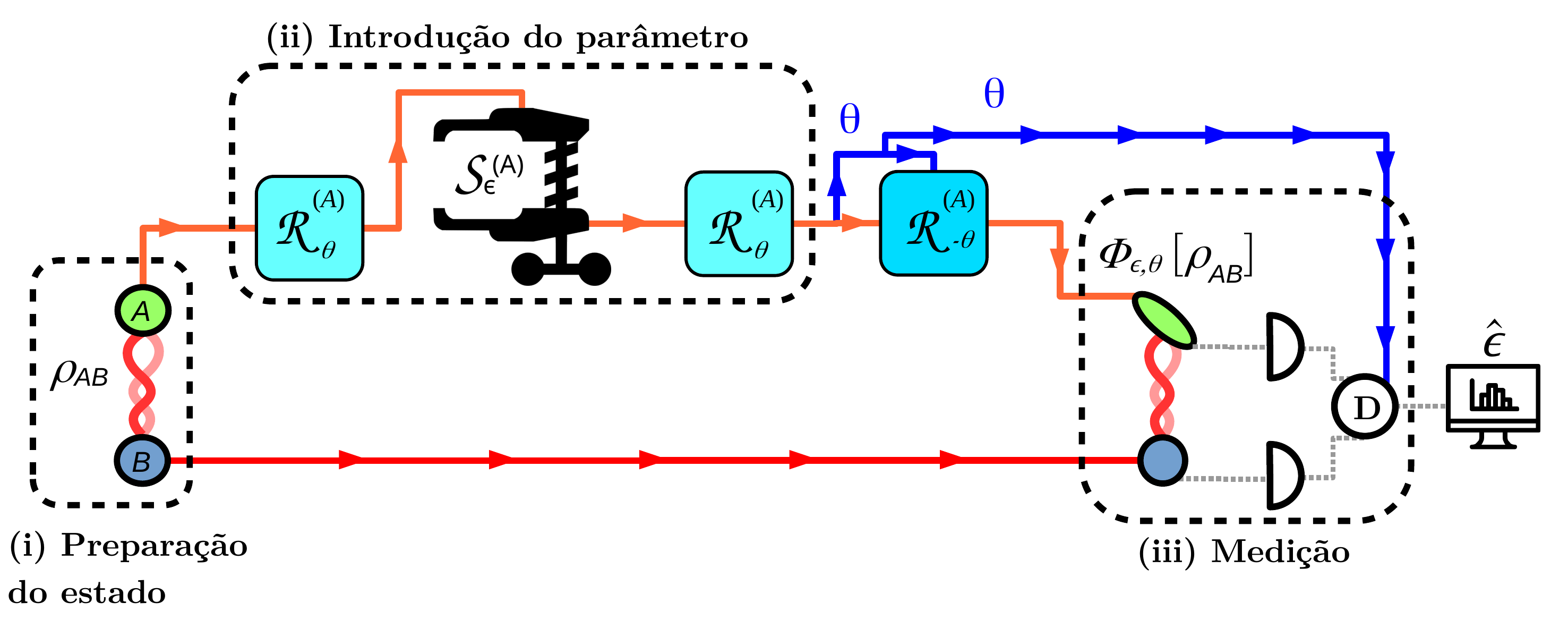}
 \end{center}
 \caption{Estratégia de Estimação Proposta. O estado é inicialmente preparado como um estado Gaussiano de um ou dois modos. O modo A é enviado a um compressor, onde o parâmetro $\epsilon$ é codificado. O modo B propaga-se livremente. Após a ação do compressor, o modo A é trazido de volta para ser medido junto ao modo B. Informações sobre a fase $\theta$ adquirida pelos modos durante o percurso da preparação até a medição são utilizadas para definir a estratégia. A Informação de Fisher Quântica média, $\overline{H_\epsilon}(\rho)$, é construída ao se fazer a média sobre todas as fases $\theta$, para estimar o parâmetro de compressão $\epsilon$. Mais detalhes são fornecidos na seção \ref{estrategia} do texto.}\label{fig1}
\end{figure}

Neste trabalho, seguimos a estratégia de estimação de \cite{Luca}. Inicialmente, o estado é preparado como um estado Gaussiano de um ou dois modos, $\rho_{AB}$. O modo A é onde o parâmetro de compressão $\epsilon$ será codificado, enquanto o modo B é uma ancila (ignorada no caso de um único modo).

Após uma primeira fase de evolução dinâmica, o \emph{squeezer} $\mathcal{S}_\epsilon^{(A)}=\exp\left(\frac{\epsilon}{2}(a^2 - (a^\dagger)^2)\right)$ atua no modo A, codificando o parâmetro $\epsilon$:

\[
\mathcal{S}_\epsilon^{(A)} [\rho_A] = U_\epsilon [\rho_A] = e^{-\frac{\epsilon}{2}(a^{\dagger 2}- a^2)} \rho_A e^{+\frac{\epsilon}{2}(a^{\dagger 2}- a^2)} = \begin{pmatrix} e^\epsilon & 0 \\ 0 & e^{-\epsilon} \end{pmatrix}.
\]

O estado final, após a segunda fase de evolução, é escrito como:

\[
(\mathcal{R}_A \mathcal{S}_\epsilon^{(A)}\mathcal{R}_A \otimes \mathcal{R}_B^2) \rho_{AB}(\mathcal{R}_A \mathcal{S}_\epsilon^{(A)}\mathcal{R}_A \otimes \mathcal{R}_B^2)^\dagger.
\]

Se $\theta_{A,B}(t)$ for conhecido, uma operação unitária $(\mathcal{R}_A^{\dagger 2} \otimes \mathcal{R}_B^{\dagger 2})$ pode remover a fase dinâmica do modo B. Assim, o mapa dinâmico total é dado por:

\[
\Phi_{\epsilon, \theta}[\rho_{AB}] = \Phi_{\epsilon, \theta}^A \otimes \mathds{1}^B [\rho_{AB}].
\]

A precisão do estimador $\hat \epsilon$ é limitada pelo limite de Cramér-Rao quântico:

\[
\delta \hat \epsilon \geq \frac{1}{\sqrt{M H_\epsilon^{(\theta)}(\rho)}},
\]

onde $H_\epsilon^{(\theta)}(\rho)$ é a Informação de Fisher Quântica (AvQFI). A versatilidade do estado pode ser avaliada pela AvQFI média (AvAvQFI):

\[
\overline{H_\epsilon}(\rho) \equiv \int_0^{2 \pi} H_\epsilon^{(\theta)}(\rho) \frac{d \theta}{2 \pi}.
\]

Com isso, $\overline{H_\epsilon}(\rho)$ define um limite inferior para o erro médio $\overline{\delta \hat \epsilon}$:

\[
\overline{\delta \hat \epsilon} \geq \frac{1}{\sqrt{M \overline{H_\epsilon}(\rho)}}.
\]

Neste artigo, estudamos $\overline{H_\epsilon}(\rho)$ para classes importantes de estados Gaussianos.

\subsection{Resultados}

Os resultados apresentados neste trabalho também constam na referência \cite{classes}. Porém, aqui tentamos apresentá-los sem tantos detalhes técnicos, e com a apresentação mais didática no início deste artigo, seguindo a apresentação na 20$^a$ Escola Matogrossense de Física.

\subsubsection{Estados monomodais}

\begin{figure}[!]
 \centering
 \includegraphics[scale=0.7]{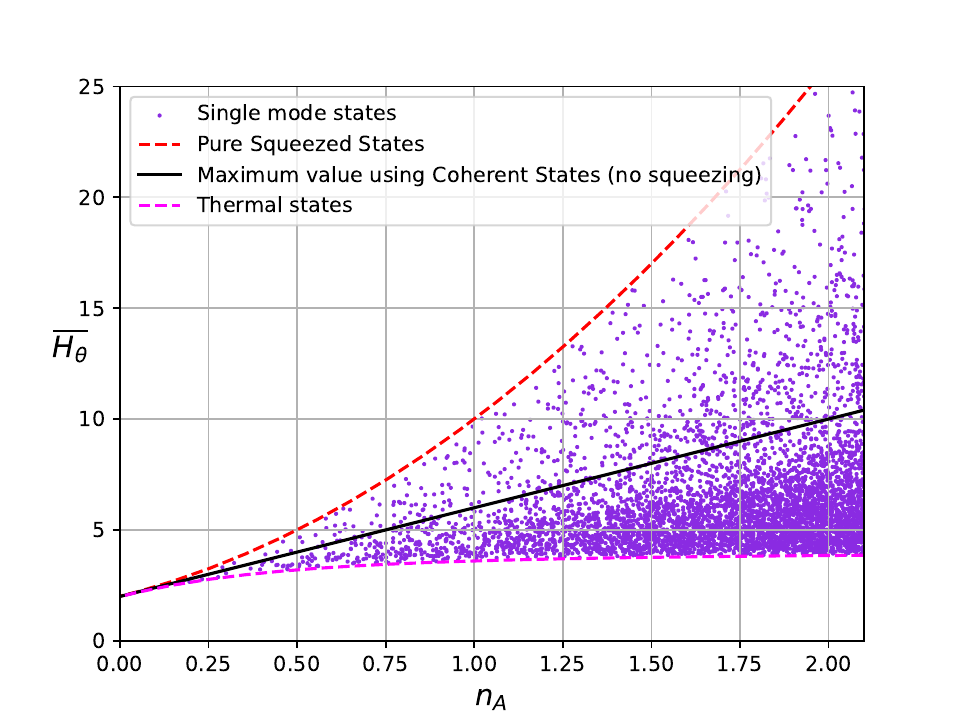}
 \caption{AvQFI média $\overline{H}_\theta$ em função do número médio de fótons $n_A$. Cada ponto representa um estado Gaussiano geral, com parâmetros aleatórios ($10^5$ estados). A linha vermelha tracejada é o limite superior para essa estratégia, obtido com estados comprimidos puros. A curva tracejada rosa representa o limite inferior, obtido por estados térmicos. A curva preta sólida mostra o valor máximo de $\overline{H}_\theta$ usando estados coerentes como estados iniciais.}
 \label{fig_single_general}
\end{figure}

Uma descrição geral do estado Gaussiano de modo único é dada pela matriz de covariância (CM):

\begin{equation}
\boldsymbol{\sigma}_{A} = \left(
\begin{array}{cc}
a & g \\
g & b
\end{array}
\right),
\label{CMgeneralsingle}
\end{equation}

com o vetor de deslocamento $\boldsymbol{\varepsilon} = (\varepsilon_x, \varepsilon_y)$, com parâmetros que representam um estado físico. Utilizando estados de modo único Gaussiano, obtemos os resultados na Figura \ref{fig_single_general}, que exibe a AvQFI média $\overline{H}_\theta$ em função do número médio de fótons $n_A$.

Os limites inferior e superior de $\overline{H}_\theta$ são dados por:

\begin{equation}
\overline{H_\theta} = 4 n_A^2 + 4 n_A + 2, \label{upper_bound_single}
\end{equation}

\begin{equation}
\overline{H_\theta} = 4 \frac{(2 n_A + 1)^2}{1 + (2 n_A + 1)^2}. \label{lower_bound_single}
\end{equation}

Para estados coerentes, a matriz de covariância é:

\begin{equation}
\boldsymbol{\sigma}_{A} = \left(
\begin{array}{cc}
a & 0 \\
0 & a
\end{array}
\right),
\label{CMgeneralcoherent}
\end{equation}

com a AvQFI média dada por:

\begin{equation}
\overline{H_\theta} = 4 n_A + 2. \label{coherent_H}
\end{equation}

\begin{figure}[!]
 \centering
 \includegraphics[scale=0.5]{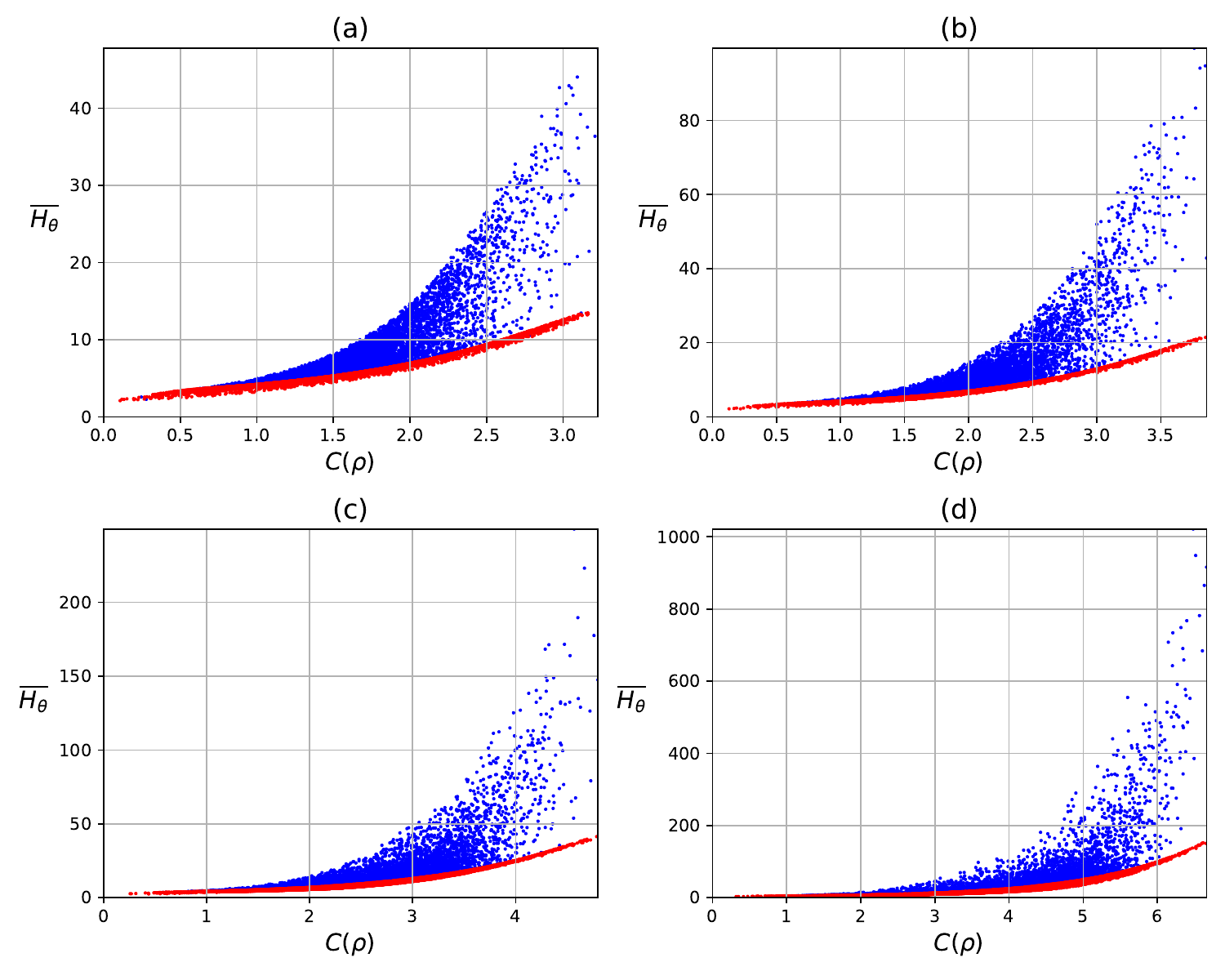}
 \caption{AvQFI média $\overline{H}_\theta$ em função da Coerência $C(\rho)$. Pontos azuis representam estados gerais de modo único e pontos vermelhos representam estados coerentes. Aumentos em $n_A$ elevam $\overline{H}_\theta$. Estados comprimidos superam estados coerentes na estimação, para o mesmo nível de Coerência.}
 \label{fig_coherence_avqfi}
\end{figure}

Na Figura \ref{fig_coherence_avqfi}, apresentamos a relação entre $\overline{H}_\theta$ e a Coerência $C(\rho)$, com estados comprimidos exibindo valores mais altos de $\overline{H}_\theta$ que os estados coerentes para o mesmo nível de coerência.

\subsubsection{Estados de dois modos}

\begin{figure}[!]
 \centering
 \includegraphics[scale=0.7]{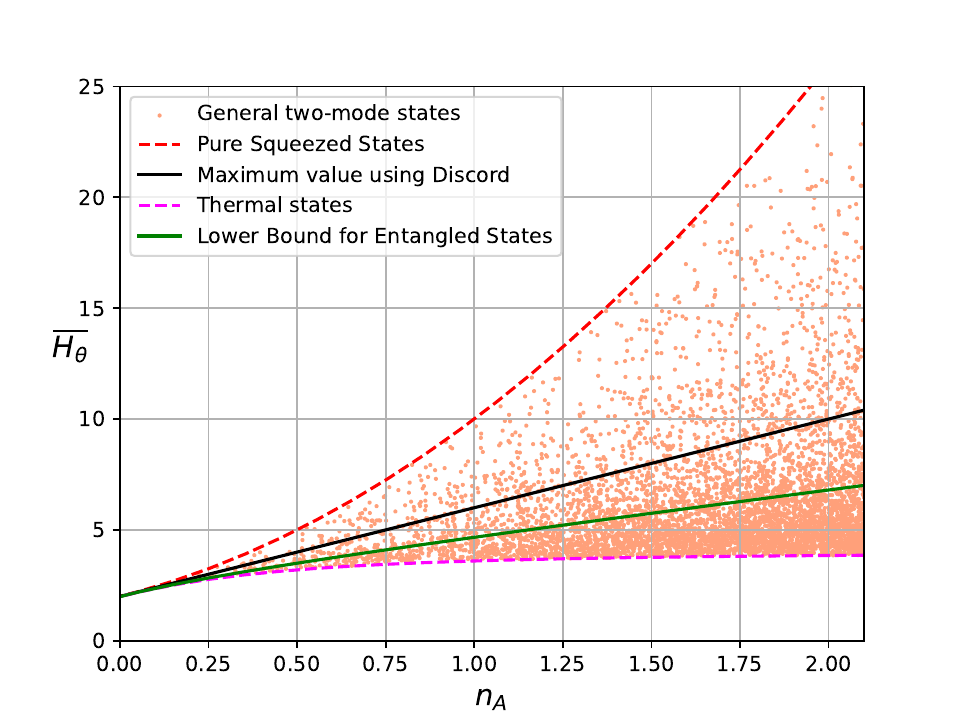}
 \caption{ IFQ médio $\overline{H}_\theta$ em função do número médio de fótons $n_A$ do modo A. Cada ponto é um estado Gaussiano de dois modos, com parâmetros de CM e vetor de deslocamento aleatoriamente escolhidos ($10^5$ estados). A linha vermelha tracejada é o limite superior para esta estratégia de estimação, retornado pelos estados comprimidos puros de dois modos. A curva rosa tracejada é o limite inferior, atingido pelos estados térmicos mistos. A curva sólida preta é o valor máximo de $\overline{H}_\theta$ usando estados contendo correlações do tipo discórdia. Finalmente, a curva verde sólida é o melhor valor possível de estimação usando estados sem correlações.}
 \label{fig_general_two_mode}
\end{figure}

Como mencionado anteriormente (seção \ref{sec_gaussian}), um estado Gaussiano de dois modos pode ser caracterizado pela seguinte CM:

\begin{equation}
\boldsymbol{\sigma}_{A B} = \left(                                                                                                                                                              \begin{array}{cccc}                                                                                                                                                               a_1 & g & c & 0 \\
g & b_1 & 0 & d \\
c & 0 & b_2 & 0 \\
0 & d & 0 & b_2
\end{array}                                                                                                                                                            \right), \label{CMgeneral2}
\end{equation}
com o vetor de deslocamento dado por: $\boldsymbol{\varepsilon} = (\varepsilon_x, \varepsilon_y, 0, 0)$. Tipicamente, os estados Gaussiano de dois modos têm o mesmo comportamento que os de um único modo. Os limites superior e inferior são dados pelos mesmos estados, os estados puros de dois modos comprimidos e os estados térmicos mistos, respectivamente, pelas equações \ref{upper_bound_single} e \ref{lower_bound_single}. A Figura \ref{fig_general_two_mode} mostra como $10^5$ estados escolhidos aleatoriamente estão distribuídos, quando estudamos $\overline{H}_\theta$ como função de $n_A$.

\subsubsection{Estados Separáveis na Forma Padrão}\label{sec_separable}

\begin{figure}[!]
 \centering
 \includegraphics[scale=0.7]{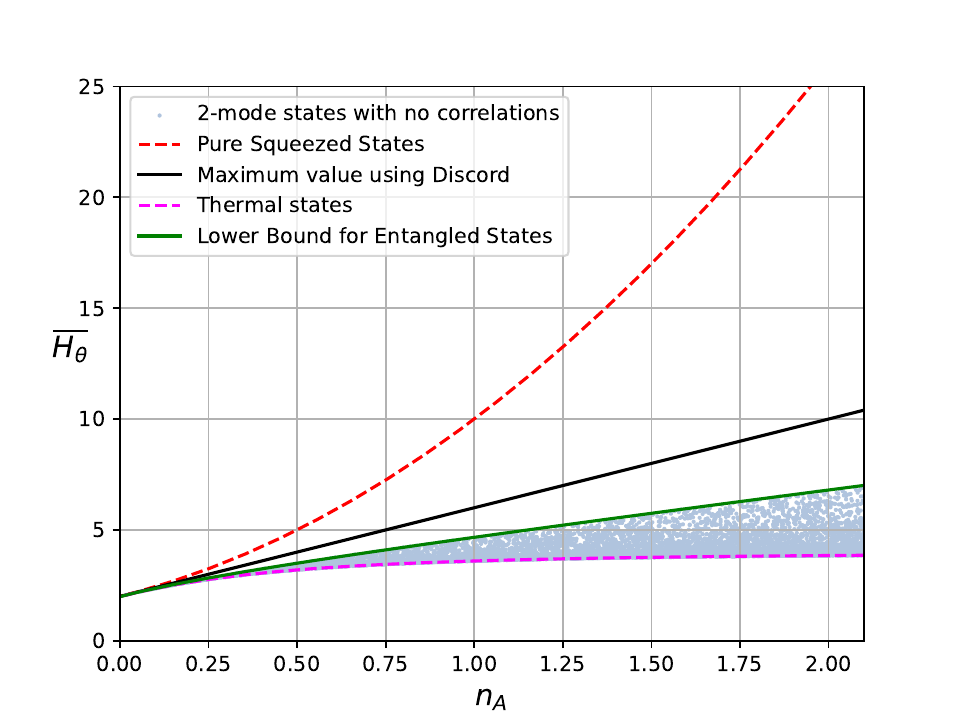}
 \caption{ IFQ médio $\overline{H}_\theta$ em função do número médio de fótons $n_A$ do modo A para estados Gaussiano separáveis. A linha verde sólida representa o valor máximo para estados sem correlações.}
 \label{fig_separable_two_mode}
\end{figure}

Começamos com o caso mais simples: estados separáveis sem correlações entre os modos A e B. Estados separáveis na forma padrão são caracterizados pela seguinte CM:

\begin{equation}
\boldsymbol{\sigma}_{A B} = \left(                                                                                                                                                              \begin{array}{cccc}                                                                                                                                                               a_1 & 0 & 0 & 0 \\
0 & b_1 & 0 & 0 \\
0 & 0 & b_2 & 0 \\
0 & 0 & 0 & b_2
\end{array}                                                                                                                                                            \right), \label{separable_standard}
\end{equation}
com vetor de deslocamento dado por: $\boldsymbol{\varepsilon} = (\varepsilon_x, \varepsilon_y, 0, 0)$. Nossos resultados para essa classe de estados são mostrados na Figura \ref{fig_separable_two_mode}. Para estados separáveis, o limite superior é dado por:

\begin{equation}
\overline{H_\theta} = 3 - \frac{1}{1 + 2 n_A} + 2 n_A. \label{upper_separable_standard}
\end{equation}

\subsubsection{Estados com Correlações do Tipo Discórdia}\label{sec_two_mode_discord}

\begin{figure}[!]
 \centering
 \includegraphics[scale=0.7]{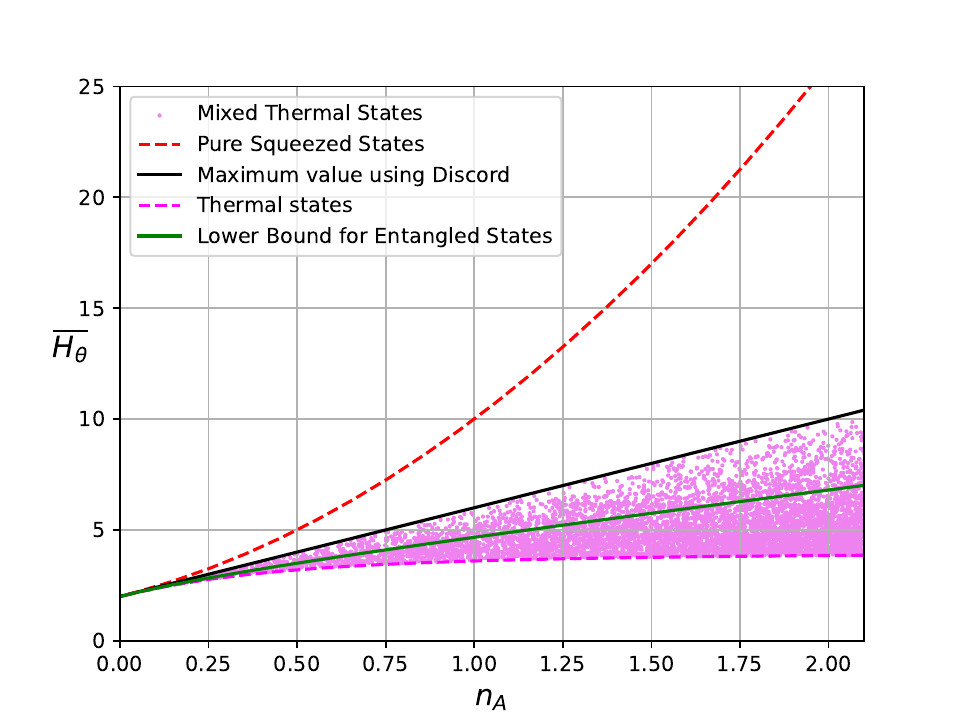}
 \caption{ IFQ médio $\overline{H}_\theta$ em função de $n_A$ para estados Gaussiano discordantes. A curva preta sólida representa o valor máximo de $\overline{H}_\theta$.}
 \label{fig_discord_two_mode}
\end{figure}

A Discórdia Quântica representa correlações quânticas em sistemas compostos que não necessariamente envolvem emaranhamento. Um estado Gaussiano discordante pode ser caracterizado pela seguinte CM:

\begin{equation}
\boldsymbol{\sigma}_{A B} = \left(                                                                                                                                                              \begin{array}{cccc}                                                                                                                                                               a & 0 & c & 0 \\
0 & a & 0 & c \\
c & 0 & b & 0 \\
0 & c & 0 & b
\end{array}                                                                                                                                                            \right), \label{discordant}
\end{equation}
com vetor de deslocamento dado por: $\boldsymbol{\varepsilon} = (\varepsilon_x, \varepsilon_y, 0, 0)$, e $c \neq 0$. Para esta classe de estados, nossos resultados são mostrados na Figura \ref{fig_discord_two_mode}, onde se observa que este tipo de estado atinge o mesmo valor máximo para $\overline{H}_\theta$ que estados coerentes de um único modo.

\subsubsection{Estados Emaranhados de Dois Modos}\label{sec_two_mode_squeezed}

Agora, analisamos os estados emaranhados puros de dois modos, com a CM dada por:

\begin{equation}
\boldsymbol{\sigma}_{A B} = \left(                                                                                                                                                              \begin{array}{cccc}                                                                                                                                                               a & 0 & c & 0 \\
0 & a & 0 & -c \\
c & 0 & b & 0 \\
0 & -c & 0 & b
\end{array}                                                                                                                                                            \right), \label{CMEntangled}
\end{equation}
com vetor de deslocamento dado por: $\boldsymbol{\varepsilon} = (0, 0, 0, 0)$. Para essa classe, o limite superior é dado por:

\begin{equation}
\overline{H_\theta} = 4 n_A^2 + 4 n_A + 2.
\end{equation}

\begin{figure}[!]
 \centering
 \includegraphics[scale=0.7]{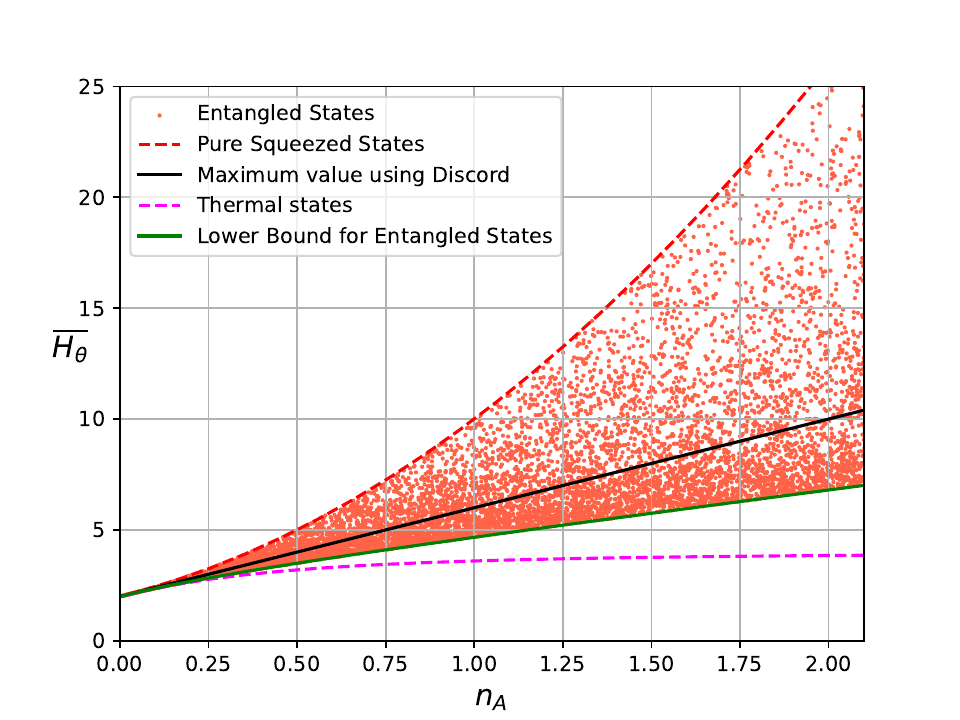}
 \caption{ IFQ médio $\overline{H}_\theta$ em função de $n_A$ para estados Gaussiano emaranhados de dois modos. A curva verde sólida representa o valor máximo para estados comprimidos puros de dois modos.}
 \label{fig_entangled}
\end{figure}

\subsubsection{Relação entre o IFQ Médio e o Emaranhamento}\label{sec_logneg}

\begin{figure}[!]
 \centering
 \includegraphics[scale=0.5]{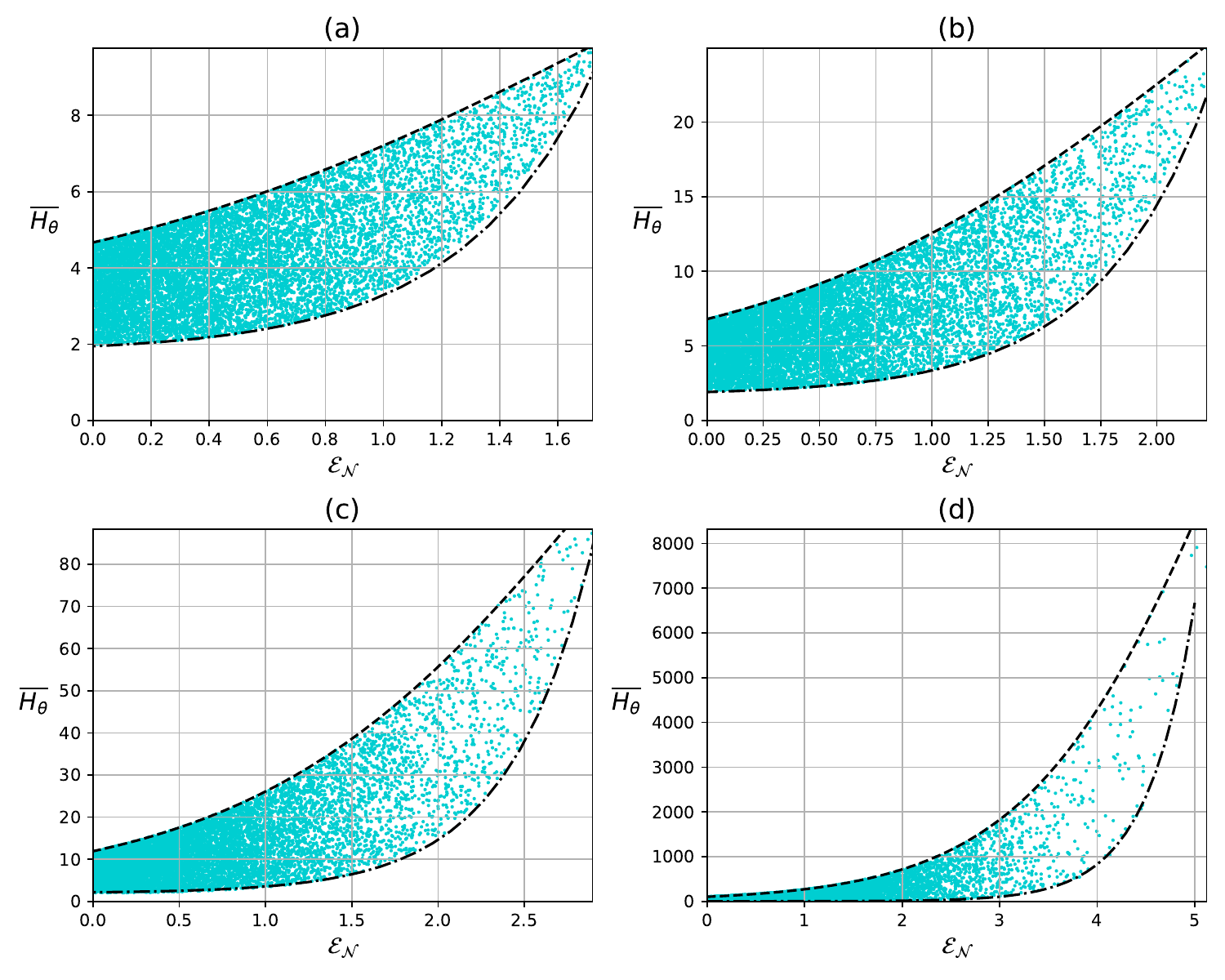}
 \caption{ IFQ médio $\overline{H}_\theta$ como função da Negatividade Logarítmica $\mathcal{E}_N$. Cada ponto representa um estado Gaussiano emaranhado de dois modos.}
 \label{logneg}
\end{figure}

Investigamos a relação entre o IFQ médio, $\overline{H}_\theta$, e o quantificador de emaranhamento, especificamente a Negatividade Logarítmica $\mathcal{E}_N$. A Figura \ref{logneg} mostra que à medida que o emaranhamento aumenta, $\overline{H}_\theta$ também aumenta. A dependência do IFQ médio em relação à Negatividade Logarítmica é dada pela seguinte equação:

\begin{equation}
\overline{H_\theta} = 2 + \frac{8n_A(1 + n_A)}{1 + (2 + 4n_A - \tilde{\nu}) \tilde{\nu}}.
\end{equation}

Para obter uma compreensão física sobre a interação entre o emaranhamento e este problema de estimação, estudamos a relação entre o IFQ Médio, $\overline{H}_\theta$, e o quantificador de emaranhamento mencionado na seção \ref{sec_gaussian}, especificamente a Negatividade Logarítmica $\mathcal{E}_N$ \cite{AdessoIP}. Primeiramente, observamos que nosso resultado depende da energia, ou seja, conseguimos obter resultados que dependem explicitamente da energia do modo A, $n_A$. Focamos nosso estudo em estados com a CM conforme descrita na equação \ref{CMEntangled}. Na Figura \ref{logneg}, mostramos nossos resultados para diferentes valores de $n_A$. O valor geral de $\overline{H}_\theta$ aumenta com $n_A$, um resultado intuitivo, já que, ao se usar um valor maior de energia, maior é a capacidade de acessar o parâmetro a ser estimado.

Na Figura \ref{logneg}, é evidente que, à medida que o emaranhamento do sistema, medido por $\mathcal{E}_N$, aumenta, o valor de $\overline{H}_\theta$ também aumenta. Um resultado importante do nosso trabalho é que o limite superior para a Figura \ref{logneg} é dado por estados puros comprimidos de dois modos, e fomos capazes de obter uma expressão analítica para essa dependência, que é dada por:

\begin{equation}
\overline{H_\theta} = 2 + \frac{8n_A(1 + n_A)}{1 + (2 + 4n_A - \tilde{\nu}) \tilde{\nu}} =  2 + \frac{8n_A(1 + n_A)}{1 + (2 + 4n_A - \mathrm{e}^{-\mathcal{E}_\mathcal{N}}) \mathrm{e}^{-\mathcal{E}_\mathcal{N}}} \label{logneq_equation},
\end{equation}

onde é clara a dependência de $\overline{H}_\theta$ em relação à medida de emaranhamento $\mathcal{E}_N$ e também à energia do modo A, $n_A$.

\section{Conclusão}\label{conclusao}

Neste trabalho apresentamos, de uma maneira didática, os princípios fundamentais da teoria estatística relacionada à estimativa de parâmetros. Mostramos como a Informação de Fisher está relacionada à precisão na estimativa de um parâmetro desconhecido introduzido no sistema, e como a Informação de Fisher pode ser interpretada à diferença de distribuições de probabilidades, quando variamos de uma pequena quantidade nosso parâmetro em questão.

Em seguida, também didaticamente, apresentamos a teoria de estimativas quântica, e também a relação entre a Informação de Fisher Quântica e o limite de Cramér-Rao quântico. A partir disso pudemos mostrar como o uso de recursos intrinsicamente quânticos podem melhorar a precisão na estimativa de parâmetros, comparando o standard quantum limit (SQL, ou limite de shot noise) com o limite de Heisenberg de precisão.

Por fim, através do estudo de estados gaussianos e suas propriedades de estimação de parâmetros, este evidenciamos as vantagens que a metrologia quântica oferece sobre a metrologia clássica, principalmente em termos de precisão nas estimativas. A análise da Informação de Fisher Quântica (IFQ) e do Limite de Cramér-Rao Quântico (QCRB) demonstrou que a presença de emaranhamento, ou compressão pura para estados monomodais, têm um impacto significativo na precisão da estimativa. A dependência da IFQ em relação à Negatividade Logarítmica ($\mathcal{E}_N$) mostra que o emaranhamento quântico não apenas aumenta a precisão das medições, mas também é um recurso crucial para ultrapassar os limites impostos pela mecânica clássica. Além disso, ao explorar o papel dos estados comprimidos e o aumento de energia no modo $A$, foi possível observar uma melhoria contínua na precisão das estimativas. Esses resultados fornecem uma base sólida para o desenvolvimento de novas técnicas em metrologia quântica e sensorística quântica, com potenciais aplicações em experimentos de alta precisão e tecnologias avançadas de medição.

\begin{acknowledgement}
O autor agradece as seguintes instituições pelo auxílio financeiro e logístico: CNPq (Conselho Nacional de Desenvolvimento Científico e Tecnológico), FAPEMIG (Fundação de Amparo à Pesquisa do Estado de Minas Gerais), FUNARBE (Fundação Arthur Bernardes), INCT-IQ (Instituto Nacional de Ciência e Tecnologia de Informação Quântica). O autor também agradece imensamente a hospitalidade de todo o Departamento de Física da Universidade Federal de Mato Grosso (UFMT) durante sua estadia para participação na 20$^a$ Escola Matogrossense de Física.
 \end{acknowledgement}


\end{document}